\documentstyle[12pt]{article}
\advance\textheight by \topskip
\begin{document}
\begin{flushright}
hep-th/9602111\\
\end{flushright}
\vspace{1cm}
\begin{center}
\baselineskip=16pt

{\Large\bf MACROSCOPIC ENTROPY OF }

{\Large\bf ${ \bf N=2}$\, EXTREMAL BLACK HOLES}  \\
\vskip 1 cm
 {\bf Andrew Strominger}\footnote{E-mail:
andy@denali.physics.ucsb.edu}\\
 \vskip 0.2cm
Department of Physics,   
University of California,
Santa Barbara,   CA
93206-9530, USA\\

\vskip 1 cm

\end{center}
\vskip 1 cm
\centerline{\bf ABSTRACT}
\begin{quotation}
Extremal BPS-saturated black holes in $N=2$, $d=4$ supergravity 
can carry electric  and magnetic charges $(q^\Lambda_{(m)},q_\Lambda^{(e)})$.
It is shown that in smooth cases the moduli fields at the horizon take a fixed 
"rational" value $X^\Lambda(q_{(m)},q^{(e)})$ which is 
determined by the charges and is independent of the asymptotic 
values of the moduli fields.
A universal formula for the 
Bekenstein-Hawking entropy is derived in terms of the charges and the 
moduli space geometry at $X^\Lambda(q_{(m)},q^{(e)})$. 
This work extends previous results of Ferrara, Kallosh 
and the author for the pure magnetic case.  
\end{quotation}
\newpage
%%%%%%%%%%%%%%%%%%%%%%%%%%%%%%%%%%%%%%%%%%%%%%%%%
\baselineskip=15pt
\section{Introduction}

The thermodynamic properties of black holes
are elegantly summarized 
by assigning a macroscopic entropy $S_{BH}$ to a black hole equal to one quarter the 
area, in Planck units, of the event horizon \cite{bek}\cite{hawk}.
Ordinary thermodynamic entropies can be derived by counting 
microstates. It is natural to suppose that this is possible for 
black hole entropy as well. However since the Planck length 
appears in the entropy formula  a quantum theory of gravity is presumably 
required. It is an outstanding challenge to string theory to 
"explain" the Bekenstein-Hawking entropy-area law. 

The simplest cases to consider are the charged, 
BPS-saturated, extremal black holes in 
an $N\ge 2$ supersymmetric theory,
which sometimes (but not always) have nonvanishing horizon area. 
These are black hole ground states: a non-extremal black hole will lose 
mass via Hawking emission 
until the BPS bound is saturated.  
$S_{BH}$ then counts the ground state degeneracy. Topological methods
can typically be employed to count such degeneracies. 
Recently such a 
microscopic derivation of $S_{BH}$ was achieved for 
some special cases in string theory \cite{ascv} - 
%\cite{cama} \cite{ghas}
\cite{spin}. However a complete explanation of why the entropy takes the 
universal form of one quarter the area in all cases remains to be found. 

In recent work by Ferrara, Kallosh and the author \cite{fks}, 
$S_{BH}$ was macroscopically computed for 
certain magnetically charged black holes in $N=2$, $d=4$ 
supergravity. A salient feature of this work was that 
the near-horizon geometry, in particular the area of the event horizon 
and values of the moduli fields,
is determined solely by the charges and is 
independent of the asymptotic values of the moduli fields $Z$. 
This comes about because the variation of the moduli fields 
as a function of the radius is governed by a damped geodesic equation on the moduli space.
The coefficients in this equation depend on the charges and the inital data 
are given by the asymptotic values of the moduli. 
The equation has an attractive fixed point $Z_q$, 
which is reached at the horizon. 
The location of this fixed point 
depends only on the charges. Hence as the asymptotic moduli are smoothly 
varied (within the basin of attraction), the near-horizon geometry is 
unaffected. As pointed out in \cite{lawi} (in the N=4 context \cite{cnsr},\cite{cvet}), this is a necessary 
condition for a 
statistical interpretation of $S_{BH}$: the number of internal 
black hole states should remain constant 
as the external environment in which it is immersed is 
adiabatically changed.  

The aim of this paper is to extend these 
results to the general case including both electric and magnetic charges. 
A simple and universal formula is derived (see section 3 below) 
which involves the moduli 
space geometry at the (charge-dependent) 
value $Z_q$ of the moduli fields at the horizon.

A microscopic derivation of this $N=2$ entropy formula 
along the lines of \cite{ascv}  
would be of interest. As mentioned in \cite{ascv}, the entropy can be 
related to the Betti numbers of moduli spaces of certain vector 
bundles on the Calabi-Yau spaces. However these numbers do not seem 
easy to compute.  It would also be of interest to see if the 
states with degeneracies defined by (\ref{vuc}) can be organized in to 
a BPS algebra of the 
kind described in \cite{jg} . The area for $N=4$ was found in 
in \cite{cnsr},\cite{cvet} and $N=8$ appeared recently in \cite{kaba}.
In all cases the entropy is independent of the moduli.

\section{Conventions and Review of Special Geometry }
We study N = 2 supergravity coupled to $n$ \ N = 2 vector multiplets. 
%framework of special geometry \cite{CREM}--\cite{CAFP}. 
In this section, 
adapted from \cite{fks},
some formulae that will be needed in the following are recalled.
Further details can be found in \cite{WLP} whose notation we adopt.
The supergravity theory is defined in terms of a projective covariantly 
holomorphic section $(X^\Lambda(\phi^i), -{i\over 2}F_\Lambda(\phi^i))$,
$\Lambda = 0, 1, ..., n, ~~i=1,...,n$, of an ~$Sp(2n+2)$ vector bundle over the
moduli space parametrized by $\phi^i$. (We note that alternate conventions
are often employed in which the  definition of $F_\Lambda$ differs by a factor
of $2i$ and/or the section is holomorphic.)
In some cases the theory can be specified by a covariantly 
holomorphic function
$F(X)$
of degree two:
\begin{equation}\label{2}
F_\Lambda(\phi^i) = F_\Lambda(X(\phi^i)) = {\partial\over
\partial X^\Lambda} F(X) \ .
\end{equation}
% Given a prepotential $F(X)$, or a holomorphic section
%$(X^\Lambda,\, -{i \over 2} F_\Lambda)$, one can construct the entire scalar
%and vector  parts of the action.
It is convenient to introduce the
inhomogeneous coordinates
\begin{equation}\label{10}
Z^\Lambda = {X^\Lambda(\phi_i)\over X^0(\phi_i)} \ , ~~~~~~
Z^0 = 1 \ .
\end{equation}
We assume $Z^\Lambda(\phi_i)$ to be invertible, so that, in special
coordinates, ${\partial Z^\Lambda\over
\partial
\phi^i}  =\delta^\Lambda_i$. ~In this case the complex scalars $Z^i = \phi^i$
{}~($
i =
1,...,n$) represent the lowest component of the $n$ vector multiplets of N = 2
supersymmetry. The K\"{a}hler potential determining the metric of these fields
is
\begin{equation}\label{22}
K(Z,\bar Z)=2\ln |X^0| = -\ln \left (N_{\Lambda\Sigma}(Z,\bar Z)\,
Z^\Lambda\bar Z^\Sigma\right ),
%&&= - \ln {1\over 2} [f(Z) + \bar f(\bar Z) + {1\over 2} (Z^i - \bar Z^i) (\bar
%f_i - f_i)] 
\end{equation}
where $N_{\Lambda\Sigma} = {1\over  4}(F_{\Lambda\Sigma} + \bar
F_{\Lambda\Sigma})$ and 
%$f(Z) = (X^0)^{-2} F(X)$.
we work in conformal gauge \cite{WLP}
\begin{equation}\label{conf}
N_{\Lambda\Sigma} X^\Lambda \bar X^{\Sigma}=1 \ .
\end{equation}
The graviphoton field strength, as well as the field strengths of the $n$
Abelian vector
multiplets, are constructed out of $n+1$ field strengths $\hat
F^\Lambda_{\mu\nu}
= \partial_\mu W^\Lambda_\nu - \partial_\nu W^\Lambda_\mu$. The graviphoton
field  strength is
\begin{equation}\label{23}
T_{\mu\nu}^{+}  = {4 N_{\Lambda\Sigma} X^\Lambda\over N_{IJ} X^I X^J} \  \hat
F_{\mu\nu}^{+\Sigma} \ ,
\end{equation}
where $\hat F^\pm\equiv \hat F \mp i *\hat F$. This combination 
of field strengths determines the central charge of the theory, since 
it enters into
the gravitino
transformation rule. The vector field strengths which enter the gaugino
supersymmetry
transformations
are
\begin{equation}\label{24}
{\cal F}_{\mu\nu}^{+\Lambda} = {\hat F}_{\mu\nu}^{+\Lambda}
-{1\over 4} X^\Lambda\, T_{\mu\nu }^{+} \ .
\end{equation}
The anti-self-dual vector field strengths ${\hat
F}^{-\Lambda}$ are part of the symplectic vector
\begin{equation}\label{11}
 \pmatrix{
{\hat F}^{-\Lambda} \cr -2i
\bar { \cal N}_{\Lambda\Sigma}\,{\hat F}^{-\Sigma} \equiv
i G^-_{\Lambda}
\cr }  ,
\end{equation}
where 
\begin{equation}\label{nln}
{\cal N}_{\Lambda\Sigma}\equiv {1 \over 4}\bar F_{\Lambda\Sigma} 
-{X^\Omega N_{\Omega\Lambda} X^\Delta N_{\Delta\Sigma}\over N_{IJ} X^I X^J}.
\end{equation}
The vector part
of the
action is then
proportional to
\begin{equation}\label{12}
\mbox{ Re}\ {\hat F}^{-\Lambda} G^-_\Lambda \ ,
\end{equation}
and the graviphoton field strength can be written in the manifestly symplectic
form
\begin{equation}\label{tst}
T_{\mu\nu}^-  =  2X^\Lambda G^-_{\Lambda\mu\nu}+ F_{\Lambda}
{\hat F}_{\mu\nu}^{-\Lambda} \ .
\end{equation}
The Lagrangian for the scalar components of the vector multiplets is
defined by
the K\"{a}hler potential as
\begin{equation}\label{13}
g_{i\bar j}\,\partial_\mu\phi^i\, \partial_\nu\bar\phi^{\bar j}\, g^{\mu\nu} \
,
\end{equation}
where $g^{\mu\nu}$ is the space-time metric, and
\begin{equation}\label{14}
g_{i\bar j}= \partial_i \partial_{\bar j} K(\phi, \bar\phi) \ .
\end{equation}

The gravitino supersymmetry transformation law, to leading order in fermi
fields, is
\begin{equation}\label{grt}\delta \psi^\alpha _\mu=2\nabla_\mu\epsilon^\alpha
-{1 \over 16}\gamma^{\nu\lambda}T^-_{\nu\lambda}\gamma_\mu
\epsilon^{\alpha \beta}
\epsilon_\beta+iA_\mu\epsilon^\alpha \ ,
\end{equation}
where $ \alpha, \beta=1,2$ are $SU(2)$ indices and  $A_\mu= {i \over 2}
N_{\Lambda\Sigma} [\bar X^\Lambda \partial_\mu  X^\Sigma
- (\partial_\mu  \bar X^\Lambda) X^\Sigma]$.
The gaugino transformation law is
\begin{equation}\label{ggt}\delta \Omega^\Lambda_\alpha
=2\gamma^\mu\nabla_\mu X^\Lambda
 \epsilon_\alpha
+{1 \over 2}\gamma^{\nu\lambda}
{\cal F}^{+\Lambda}_{\nu\lambda} \epsilon_{\alpha \beta}
\epsilon^\beta+2i\gamma^\mu A_\mu\epsilon_\alpha \ .
 \end{equation}

BPS states of the N = 2 theory have a mass equal to the central charge $z$.
It follows from the supersymmetry transformation rules that this is simply
the graviphoton charge
\cite{CAFP}
\begin{equation}\label{15}
M = |z|  =
  |q^{(e)}_\Lambda X^\Lambda - {i \over 2} q_{(m)}^\Lambda F_\Lambda|
%=|q^{(e)}_0 +  q^{(e)}_i Z^i + {i \over 2} (q_{(m)}^0 Z^i -
%q_{(m)}^i)f_i - iq^0_{(m)} f|
\ ,
\end{equation}
where the charges
\begin{eqnarray}\label{15nn}
q^\Lambda_{(m)}&&\equiv {1 \over 2 \pi}\int_{S^2}{\rm Re}\hat F^{+\Lambda},\nonumber \\
~~~~~~~~~~~~~q_\Lambda^{(e)}&&\equiv-{1 \over 2 \pi}\int_{S^2}{\rm Re}
(iG^+_\Lambda),
\end{eqnarray}
comprise a symplectic vector. Quantum mechanically these charges are quantized. 
The minimal integral normalization of (\ref{15nn}) should be determined in the 
context of a complete quantum theory such as string theory.

\section{Black Hole Entropy}

In \cite{fks} pure magnetically charged extremal black holes 
were described (in theories with restricted prepotentials ) 
as solitons which interpolate, as one moves 
into the throat of the black hole,  between 
the asymptotic Minkowski vacuum and a maximally symmetric charged 
Robinson-Berttoti ($AdS \times S^2$) vacuum 
at the horizon. The Minkowski vacuum is characterized by a fixed value of the 
moduli $Z^\Lambda_\infty$, while the Robinson-Berttoti 
vacuum is characterized 
by a fixed value of the 
moduli 
\begin{equation}\label{aa}
Z^\Lambda_q=q^\Lambda_{(m)}/q^0_{(m)}~~~{\rm when}~~q_\Lambda^{(e)}=0
\end{equation} 
determined solely by the charges. 
The variation of the moduli $Z^\Lambda$ as a function of the radius 
is given by a (charge-dependent) damped geodesic equation which has an 
attractive fixed point at  
$Z^\Lambda_q$. Hence as long as one remains in the basin of attraction, 
the near horizon geometry depends only on the charges and 
is unaffected by smooth variations of $Z^\Lambda_\infty$. 
The area of the event horizon is given by the universal 
formula 
\begin{equation}\label{bb}
Area(q) ={\pi \over 4} (q^0_{(m)})^2e^{-K(Z_q)}.
\end{equation} 
This may alternately be expressed in the covariant form
\begin{equation}\label{bbb}
Area(q) ={\pi \over 4} q^\Lambda_{(m)}N_{\Lambda \Sigma}q^\Sigma_{(m)}.
\end{equation} 
For the special case 
\begin{equation}\label{aabb}
Z^\Lambda_\infty =Z^\Lambda_q,
\end{equation} 
the geometry is Reissner-Nordstrom 
\begin{equation}\label{17}
ds^2 = -e^{2U} dt^2 + e^{-2U} (dr^2 +r^2d\Omega^2)\ 
\end{equation}
with
\begin{equation}\label{uc}
e^{-U}=1+{ \sqrt{Area(q)/4\pi}\over  r}\,  \ .
\end{equation}

We wish to generalize (\ref{aa}) and (\ref{bb}) to the general case in 
which both electric and magnetic charges are nonzero and the 
prepotential $F$ is unrestricted. In \cite{fks} 
the full spacetime solutions were 
described for $Z^\Lambda_\infty \neq Z^\Lambda_q$. 
This seems rather difficult 
to accomplish for the general case. However in order to 
generalize (\ref{aa}) and (\ref{bb}) we need only understand the 
special case of constant $Z^\Lambda$, which is a much simpler problem. 
Under these 
circumstances the gaugino  variation reduces to 
\begin{equation}\label{ggtr}\delta \Omega^\Lambda_\alpha
={1 \over 2}\gamma^{\nu\lambda}
{\cal F}^{+\Lambda}_{\nu\lambda} \epsilon_{\alpha \beta}
\epsilon^\beta \ .
 \end{equation}
Multiplying this equation by $\gamma^{\nu\lambda}
{\cal F}^{+\Lambda}_{\nu\lambda}$ 
one finds that the right hand side can vanish 
if and only if 
\begin{equation}\label{gtr}
{\cal F}^{+\Lambda}_{\nu\lambda}=0.
\end{equation}
It follows from the definition (\ref{24}) of $\cal F^+$ in 
terms of $\hat F^+$ that at 
regular points in the moduli space solutions of this equation are 
parameterized by a complex spacetime constant $C$ 
\begin{equation}\label{ccc}
{\hat F}^{+\Lambda}_{\nu\lambda}=C X^\Lambda \epsilon^+_{\nu\lambda}
\end{equation} 
where $\epsilon^+$ obeys $*\epsilon^+=i\epsilon^+$ and is normalized 
so that its integral over the nontrivial $S^2$ equals $2\pi$. In 
general the phase of $C$ depends on the Kahler gauge. 
Inserting this in to the definitions (\ref{11}) and (\ref{nln}) one finds 
\begin{equation}\label{gqn}
iG_\Lambda^+=C{i \over 2}F_\Lambda \epsilon^+ ~.
\end{equation}
The electric and magnetic 
charges are accordingly 
\begin{equation}\label{ddd}
(q_{(m)}^\Lambda ,q^{(e)}_\Lambda  )={\rm Re }(C X^\Lambda_q,-C{i \over 2}
F_\Lambda(X_q) ) .
\end{equation}
If $\Lambda=0,1,...n$, the right hand side of this equation involves  
$2n+4$ variables, the real and imaginary parts of $X^\Lambda, C$. Two 
of these may be eliminated by 
a choice of Kahler gauge leaving $2n+2$ real variables. Hence 
(\ref{ddd}) is $2n+2$ equations for $2n+2$ unknowns, and both $X^\Lambda$ 
and $C$ are determined up to gauge transformations.
A solution of this equation associates a point on moduli space, 
which we shall 
denote $Z^\Lambda_q=X^\Lambda_q/X^0_q$, to a  
quantized set of charges 
$(q_{(m)}^\Lambda , q^{(e)}_\Lambda )$.(In principle there 
could be more or less than one solution.) This point is invariant under 
uniform rescalings of all the charges, while $C$ scales linearly.  

Given (\ref{ccc}) the vanishing of the gravitino variation determines the 
metric. One finds, as in \cite{fks}, that the geometry is of the form (\ref{17})
with
\begin{equation}\label{ubc}
e^{-U}=1+{\sqrt{C\bar C}\over 4 r}\,  \ .
\end{equation}
This corresponds to a Reissner-Nordstrom spacetime with 
horizon area 
\begin{equation}\label{vuc}
Area(q_{(e)}, q^{(m)}) ={\pi \over 4} C\bar C.
\end{equation}
When $F$ is real for 
real $X^\Lambda$ 
and $q^{(e)}=0$, then $C X^\Lambda_q = q_{(m)}^\Lambda$, 
and $C\bar C=(q^0_{(m)})^2e^{-K(Z_q)}$. This is in 
agreement with \cite{fks} and equations (\ref{aa}) and (\ref{bb}).

The relation 
(\ref{ddd}) can be satisfied for quantized $(q^{(e)}, q_{(m)})$ 
only at special rational points on moduli space, 
the analogs of rational tori. Insight 
into the nature of these points 
can be gained in the context of string compactification on a Calabi-Yau 
space. 
The holomorphic 3-form on the Calabi-Yau can be expanded 
\begin{equation}\label{eee}
\Omega=e^{-K/2}X^\Lambda \alpha_\Lambda -{i \over 2}
e^{-K/2}F_\Lambda \beta^\Lambda ,
\end{equation}
where $\alpha_\Lambda$ and $  \beta^\Lambda$ are a symplectic integral basis 
for $H_3$ obeying
\begin{equation}\label{vuvc}
\int \alpha_\Lambda \wedge \beta^\Sigma={\delta_\Lambda}^\Sigma . 
\end{equation}
This implies 
${\rm Re}[ C e^{K/2} \Omega ]$ is proportional to an element of $H_3(X,Z)$ :
\begin{equation}\label{vcc}
{\rm Re}[ C e^{K/2} \Omega ]=q_{(m)}^\Lambda\alpha_\Lambda + q^{(e)}_\Lambda \beta^\Lambda. 
\end{equation}
Typically one specifies half the $b_3$ complex 
periods of $\Omega$, $e^{-K/2}X^\Lambda$, and 
then computes the second half, $e^{-K/2}F_\Lambda$, as a 
function of the first half.  
Here the situation is different: the 
point on moduli space corresponding to a given set of charges is determined  
by (in a gauge in which $C$ is real) the real part of all $b_3$ periods. 
In general it is a difficult problem to solve for the complex $X$'s 
as a function of the real parts of the periods, and a solution may not exist for 
all values of the $(q^{(e)}, q_{(m)})$. (Extremal black hole 
solutions with such charges will have a different character and 
in some cases have a naked singularity.) In some simple cases the 
solutions can be explicitly found \cite{fks}.

%The area formula (\ref{vuc}) also applies to the case 
%where the moduli arise from Kahler deformations.
%At large radius (\ref{vuc}) becomes 
%\begin{equation}\label{vucl}
%Area(q_{(m)}, q^{(e)}) ={32 \over \pi} (q^0_{(m)})^2V_6,
%\end{equation}
%where $V_6$ is the volume of the Calabi-Yau at $Z_q$. This linear volume 
%dependence suggests that 
%the entropy has a microscopic origin in degrees of freedom moving on the 
%Calabi-Yau. 

\section*{Acknowledgements}
This work was supported by DOE grant DOE-91ER40618. 
We are grateful to S. Ferrara, R. Kallosh and C. Vafa for useful
conversations.

%\newpage

\end{document}